\documentclass[sigconf]{acmart}

\usepackage{multirow}
\usepackage{multicol}

\renewcommand\footnotetextcopyrightpermission[1]{} % removes footnote with conference information in first column
\AtBeginDocument{%
  \providecommand\BibTeX{{%
    \normalfont B\kern-0.5em{\scshape i\kern-0.25em b}\kern-0.8em\TeX}}}

\setcopyright{acmcopyright}
\copyrightyear{2018}
\acmYear{2018}
\acmDOI{10.1145/1122445.1122456}

\acmConference[]{}{}{}
\acmBooktitle{Woodstock '18: ACM Symposium on Neural Gaze Detection,
  June 03--05, 2018, Woodstock, NY}
\acmPrice{15.00}
\acmISBN{978-1-4503-XXXX-X/18/06}

\begin{document}

% \title{Co-BERT: An End-to-End Pseudo Relevance Feedback Model using BERT}
\title{Co-BERT: A Context-Aware BERT Retrieval Model Incorporating Local and Query-specific Context}
%
% The "author" command and its associated commands are used to define
% the authors and their affiliations.
% Of note is the shared affiliation of the first two authors, and the
% "authornote" and "authornotemark" commands
% used to denote shared contribution to the research.
% \affiliation{%

\author{Xiaoyang Chen}
\affiliation{%
  \institution{University of Chinese Academy of Sciences}
  \city{Beijing}
  \country{China}}
\email{chenxiaoyang19@mails.ucas.ac.cn}

\author{Kai Hui}
\authornote{This work was done before joining Amazon.}
\affiliation{%
  \institution{Amazon Alexa AI}
  \city{Berlin}
  \country{Germany}
}
\email{kaihuibj@amazon.com }

\author{Ben He}
\affiliation{%
  \institution{University of Chinese Academy of Sciences}
  \city{Beijing}
  \country{China}
}
\email{benhe@ucas.ac.cn}

\author{Xianpei Han}
\affiliation{%
  \institution{Institute of Software, Chinese Academy of Sciences}
  \city{Beijing}
  \country{China}
}
\email{xianpei@iscas.ac.cn}

\author{Le Sun}
\affiliation{%
  \institution{Institute of Software, Chinese Academy of Sciences}
  \city{Beijing}
  \country{China}
}
\email{sunle@iscas.ac.cn}

\author{Zheng Ye}
\affiliation{%
  \institution{South-Central University For Nationalities}
  \city{Hubei}
  \country{China}
}
\email{yezheng@scuec.edu.cn}

% \affiliation{%

% \affiliation{%

% \affiliation{%

\renewcommand{\shortauthors}{Anonymous}

\begin{abstract}
 BERT-based text ranking models have dramatically advanced the state-of-the-art
  in ad-hoc retrieval, wherein most models tend to consider individual query-document pairs 
  independently. In the meantime, the importance and usefulness 
  to consider the cross-documents interactions and the
  query-specific characteristics in a ranking model have been repeatedly confirmed, mostly in the 
  context of learning to rank.
  The BERT-based ranking model, however, has not been able to fully incorporate 
  these two types of ranking context, thereby ignoring the inter-document relationships from the ranking
  and the differences among queries.
  To mitigate this gap, 
  in this work, an end-to-end transformer-based ranking model, named Co-BERT, has been proposed 
  to exploit several BERT architectures to calibrate the query-document representations using
  pseudo relevance feedback before modeling the relevance of a group of documents jointly. 
  Extensive experiments on two standard test collections confirm the 
  effectiveness of the proposed model in improving the performance of text re-ranking
  over strong fine-tuned BERT-Base baselines.
  We plan to make our implementation open source to enable further comparisons. 
\end{abstract}

\maketitle

\section{Introduction}\label{sec.introduction}
Recent advances in information retrieval (IR) models have shown promising performance gain on ad-hoc text retrieval tasks by utilizing large-scale pre-trained transformer-based language models, e.g. Bidirectional Encoder Representations from Transformers (BERT)~\cite{DBLP:conf/naacl/DevlinCLT19}, improving upon classical IR models by a wide margin on different benchmarks~\cite{DBLP:journals/corr/abs-1901-04085,DBLP:conf/sigir/DaiC19,DBLP:conf/emnlp/YilmazYZL19,DBLP:journals/corr/abs-2008-09093}.

% learn groupwise/document interaction is important
Most of these existing BERT-based ranking models, however,
consider query-document pairs independently, 
following the premise of  probability ranking principle (PRP)~\cite{robertson1977probability}.
A well-known observation indicates that,
unlike in ordinal classification, 
the main goal of a ranking problem 
is to optimize ranking lists given queries, making 
the consideration of the context of the ranking 
important, including the \textit{local ranking context} in terms of cross-document interactions~\cite{DBLP:conf/ictir/AiWBGBN19,DBLP:journals/corr/abs-1910-09676,DBLP:conf/sigir/PangXALCW20}
and the \textit{query-specific context} incorporating different characteristics among different queries~\cite{DBLP:conf/ictir/AiWBGBN19}.
There have been many successful attempts to incorporate either
context, mostly in learning to rank, wherein the handcrafted 
features serve as query-document representations. 
In early works, abundant loss functions have been proposed to
optimize on top of a pair or a list of documents in the context of
learning to rank~\cite{liu2010learning, li2011short},
modeling the cross-document interactions at loss level,
achieving superior performance on L2R benchmark~\cite{qin2010letor}.
In addition, 
a groupwise ranking framework for multivariate scoring functions is proposed~\cite{DBLP:conf/ictir/AiWBGBN19} to
determine  the relevance scores of a group of documents jointly,
taking handcrafted learning-to-rank features as query-document presentations and
using stack of dense layers to evaluate the relevance. 
More recently,
a neural learning-to-rank model named SetRank is proposed to
directly learn a ranking model defined on document sets, employing 
a stack of multi-head self-attention blocks to learn the embedding for all documents jointly,
successfully incorporating the local context. 
As for the query-specific context, 
Ai et. al.~\cite{DBLP:conf/sigir/AiBGC18} pointed out that
relevant documents for different queries may have different distributions
and proposed Deep Listwise Context Model to exploit 
the pseudo relevance feedback documents to
inject information about the query's characteristics
into the learning to rank scorer, making the model aware of the query-specific distributions.

To the best of our knowledge, however,
none of these existing works has been able to encode both
local ranking context and query-specific context into a BERT-based ranking architecture,
combining the valuable observations and the powerful pre-trained models. 
Actually,
as mentioned in~\cite{Qiao2019UnderstandingTB},
using pairwise loss when employing BERT for ranking
does not lead to improvements, suggesting that
it is non-trivial to directly reuse the pairwise loss
together with the BERT-based ranker.
For the modeling beyond single query-document pairs using BERT, 
duoBERT~\cite{DBLP:journals/corr/abs-1910-14424}
concatenates two documents and the query together before feeding into BERT layers, and the
output from BERT is trained to learn a pairwise comparisons between two documents. 
However, 
there exist no straightforward extension to incorporate 
the full local ranking context using duoBERT 
due to the facts that BERT model can not 
encode sequence which is too long.
In addition, duoBERT does not consider the query-specific context in the design.
In facts, we are not aware of any existing works that attempt to 
incorporate the query-specific context into a BERT-based ranker. 
The closest efforts are the recent attempts to incorporate the pseudo relevance feedback (PRF) into BERT model,
which, however, are motivated to mitigate the representation gaps between 
queries and documents as in~\cite{padaki2020rethinking,DBLP:journals/corr/abs-2101-07918, zheng2020bert}.
In a nutshell, both local ranking context and the query-specific context have been
shown effective in learning to rank and non-BERT neural IR models.
But they are yet to be realised for BERT-based ranking model,
since it is challenging to 
incorporate the powerful transformer architecture, 
the strength of the pre-trained language model,
and the benefits of the context at
the same time without introducing too much computation overheads.

% our methods
In aware of this gap,
in this work, we propose a context-aware BERT-based ranking model, coined as Co-BERT,
which is equipped to consider both local ranking context and the 
query-specific context during scoring.
The proposed Co-BERT includes two components, namely,
the pseudo relevance feedback (PRF) calibrator that incorporates the query-specific context,
and groupwise scorer which  
captures the local ranking context among different candidate documents.
In particular,
in the PRF calibrator,
given a query,
several top-ranked PRF documents are selected,
and their query-document interaction presentation from BERT 
are used as the prototypes.
Given a document for evaluation, 
the query and the document are first concatenated and passed through multiple BERT layers.
The interaction representation from the output layer 
% (namely, the
% output representation of $[CLS]$ token from BERT)
is used to encode the interaction between them.
Instead of directly using such interaction presentation,
the prototypes based on PRF are used to 
calibrate the interaction representations to incorporate the query-specific context, e.g., the query difficulty, 
using a two-layers BERT.
In the groupwise scorer, 
inspired by~\cite{DBLP:conf/sigir/PangXALCW20},
$n$ candidate documents are grouped together and their calibrated interaction representations are 
passed through several BERT layers 
for the modeling of the local ranking context,
before projecting the outputs into $n$ ranking scores.
The Co-BERT model is trained end-to-end using pointwise loss function.

To this end, a novel context-aware ranking model, coined as Co-BERT, 
is proposed to incorporate both local ranking and query-specific context into
the BERT-based ranking model using three BERT architectures, exploiting the 
strength of the pre-trained models in both PRF calibrator and interaction encoder.
Put together, our contributions in this paper are threefold.
\begin{enumerate}
    \item 
    To the best of our knowledge, this is the first successful attempt 
    to incorporate both local ranking and query-specific context into a BERT-based ranking model, 
    ending up with a novel ranking model named Co-BERT.
    
    \item Evaluation on two standard TREC test collections, namely, Robust04 and GOV2, 
    demonstrates that the proposed Co-BERT could advance the state-of-the-art BERT-based ranking model
    by a big margin. For example, Co-BERT boosts nDCG@20 and MAP@1K by 14.7\% and 13.9\%, 
    respectively on GOV2, compared with the state-of-the-art BERT-Base ranking model which has been
    pre-trained on MS Marco~\cite{DBLP:journals/corr/abs-1901-04085} before fine-tuning.

    \item We also demonstrate that, while providing a big improvements,
    the computation overhead required by Co-BERT during inference is also remarkable, where
    Co-BERT requires as less as 2.4\% extra computations compared with
    a standard BERT-Base ranking model on GOV2.
\end{enumerate}

\section{Related Work}\label{sec.relatedwork}

 \subsection{BERT-based ranking models}

The contextualized pre-trained language model, namely Bidirectional Encoder Representations from Transformers (BERT), has shown to outperform the state-of-the-art in many natural language processing tasks including document ranking~\cite{DBLP:conf/naacl/DevlinCLT19}. 
More recently, Nogueira and Cho~\cite{DBLP:journals/corr/abs-1901-04085} utilize the MS MARCO~\cite{nguyen2016ms} and TREC-CAR~\cite{dietz2017trec} datasets with large amount of training samples to train BERT passage re-rankers, and demonstrate significantly improved retrieval performance over unsupervised baselines and the existing shallow ranking models.
To overcome BERT's 512 limit of sequence length, Dai and Callan~\cite{DBLP:conf/sigir/DaiC19} split a document into fixed length passages and use a BERT re-ranker to predict the relevance of each passage independently. 
Beyond that, 
MacAvaney et al.~\cite{DBLP:conf/sigir/MacAvaneyYCG19} incorporate BERT's classification vector into existing neural models;
Yilmaz et al.~\cite{DBLP:conf/emnlp/YilmazYZL19} utilize larger datasets with more training samples by transferring models across different domains and aggregate sentence-level evidences to rank documents; 
Nogueira et al.~\cite{DBLP:journals/corr/abs-1910-14424} propose a multi-stage ranking architecture with BERT that can trade-off quality against latency;
 Wu et al.~\cite{DBLP:conf/www/WuMLZZZM20} propose the context-aware Passage-level Cumulative Gain to aggregate passage relevance representations scores, which is incorporated into a BERT-based model for document ranking.
 As can be seen, apart from~\cite{DBLP:journals/corr/abs-1910-14424}, where a pairwise relativity has been modeled,
the mentioned BERT-based ranking models consider query-document pairs independently, and
ignore both cross-document interactions and query-specific differences.

 There are also works that exploit pseudo relevance feedback information to boost ranking. 
Padaki et al.~\cite{padaki2020rethinking} investigate several traditional keyword expansion approaches and find that they are not necessarily beneficial to improving BERT's ranking performance. Thereafter, Zheng et al.~\cite{zheng2020bert} propose BERT-QE that expands the original query by text snippets, instead of individual keywords, selected by a fine-tuned BERT ranker. 
Very recently, Yu et al.~\cite{DBLP:journals/corr/abs-2101-07918} propose PGT, a transformer-based pseudo relevance feedback approach. In PGT, pseudo relevance feedback is performed by feeding a concatenation of each feedback document with the target document into BERT. The resulting [CLS] tokens by using different feedback documents are combined within a graph-based transformer model for the final relevance weighting. 
Though PRF have been incorporated into BERT in these works,
unlike Co-BERT,
the motivation is to expand the queries to mitigate the vocabulary mismatch between the queries and the documents instead of 
deriving prototypes serving as query-specific context. Inspired by~\cite{DBLP:conf/sigir/AiBGC18}, 
Co-BERT attempts to exploit the interactions of the PRF documents as the prototypes to
inject the query-specific information into the model.

\subsection{Local Ranking and Query-specific Context}
The nature of the ranking problem 
makes the optimizations on top of ranking lists important,  
and the models should consider other documents when evaluating a single query-document pair.
In early works, pairwise and listwise loss functions have been proposed 
to learn from a pair or a list of documents~\cite{liu2010learning, li2011short},
beyond the pointwise loss wherein query-document pairs are considered independently. 
More recently,
the cross-document interactions have been further incorporated into the ranking models.
Ai et al.~\cite{DBLP:conf/ictir/AiWBGBN19} propose a general framework for multivariate scoring functions, in which the relevance score of a document is determined by taking multiple other documents in the list into consideration, instead of the traditional pointwise document scoring. Pang et al.~\cite{DBLP:conf/sigir/PangXALCW20} propose a neural learning-to-rank approach, named SetRank, that directly learns a permutation-invariant ranking model defined on document sets of unlimited size. Evaluation on learning to rank datasets shows performance gain of SetRank over strong baselines. 
Pasumarthi et al. \cite{DBLP:journals/corr/abs-1910-09676} leverage the cross-document interaction by a self-attention based neural network, demonstrating improved effectiveness and efficiency on several learning to rank datasets.
In addition, Ai et al.~\cite{DBLP:conf/sigir/AiBGC18} 
argue that relevant documents for different queries often have different distributions in feature space, and 
suggest that a ranking model should take into account the query-specific feature distributions.
Accordingly, Ai et al.~\cite{DBLP:conf/sigir/AiBGC18}
propose to employ a recurrent neural network to encode the top-ranked results, from which a context model is learned to
incorporate the query-specific feature distributions. 

On the one hand, the convincing improvements reported in these existing works highlight the importance and the 
usefulness for a ranking model to consider the cross-documents interactions, which is named ``local ranking context'' in this work,
and the query-specific distributions, which is coined ``query-specific context'' here.
On the other hand, to the best of our knowledge, these existing models are built upon 
handcrafted features and do not have straightforward extensions to make uses of the more recent BERT-based
ranking models and the pre-trained language models, which have advanced the state-of-the-art by a wide margin.
Moreover, none of these mentioned model has considered both kinds of context, and they always focus on 
one or another type of context during the modeling. 
Complementary to these models, Co-BERT successfully incorporates both local ranking and query-specific context into a BERT-based ranking model.
\section{Method}\label{sec.method}

\begin{figure*}
\centering
\includegraphics[width=0.95\textwidth]{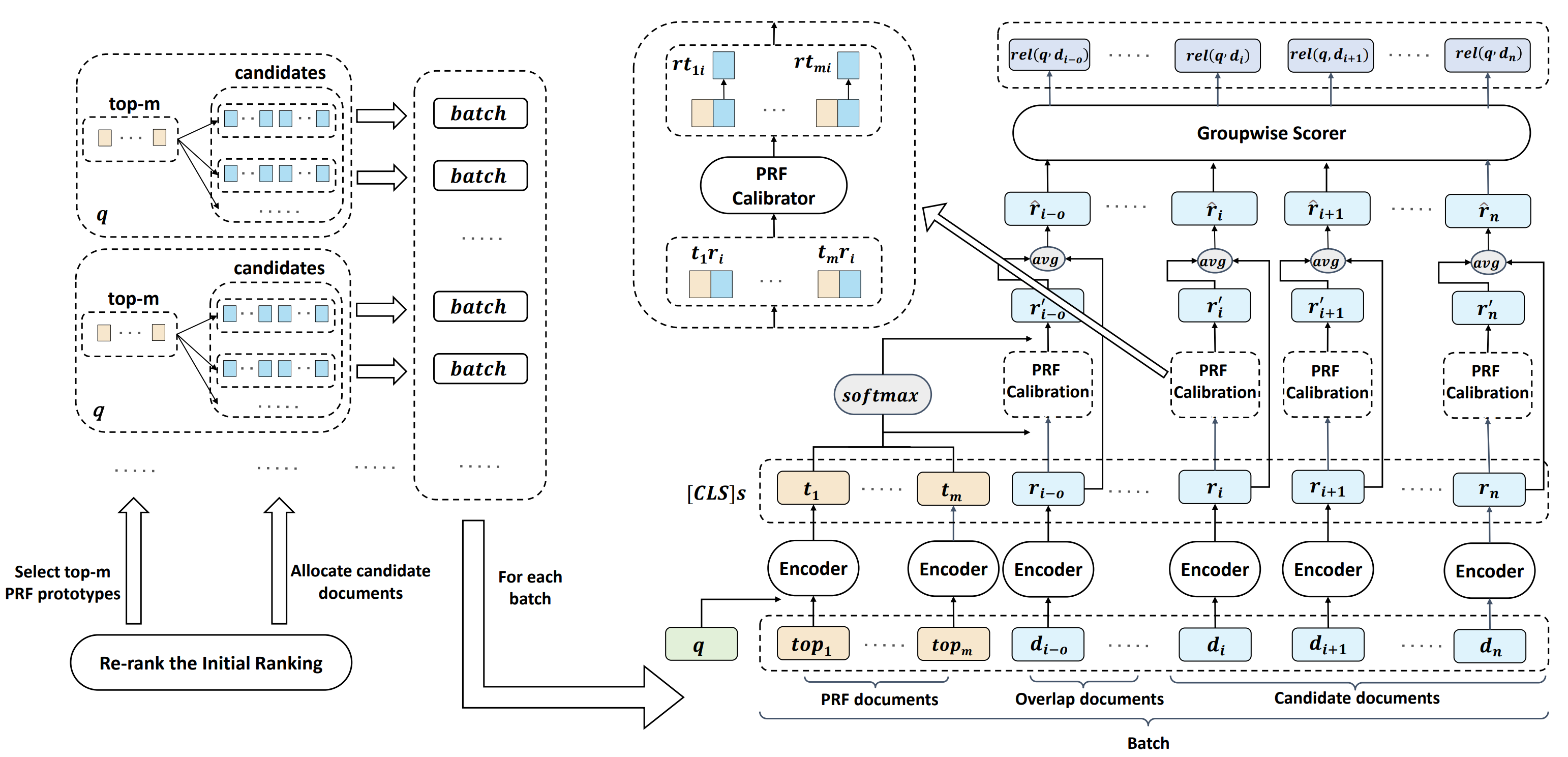}
\caption{Model architecture of the proposed Co-BERT framework. During training, documents in the initial ranking are allocated into batches. Each batch contains the top-{\textit m} documents selected by a BERT ranker as relevance prototypes, a small set of the last {\textit o} documents from the previous batch in order to maintain the cross-references between batches, and a sequence of candidate documents. Each candidate document is encoded by interaction with the query, and subsequently with the PRF documents representations, before being fed into the groupwise scorer. Batches from different queries are randomly fed for training. During inference, the framework follows the same procedure by assigning relevance scores for batches of candidate documents on a per-query basis. }
\label{fig.model_architecture}
\end{figure*}

In this section, we present the proposed Co-BERT method for document re-ranking, 
which could be trained
end-to-end, and could boost the ranking quality significantly by the calibration using
the pseudo relevance feedback documents and the groupwise scorer to incorporate cross-documents interactions.  
The model architecture is summarized in Figure~\ref{fig.model_architecture}. 

\subsection{Overview}\label{sec.overview}
Given a query and $k$ ranked documents, e.g., from BM25, 
an end-to-end framework is proposed to boost the ranking of these $k$ documents.
As discussed, referring to other candidate documents from the same ranking is important.
The proposed Co-BERT includes two components for this: 
Calibration using pseudo relevance feedback and the cross-documents interaction encoder named groupwise scorer. 
In particular, during the calibration, 
the top-$m$ from the input $k$ documents are selected as pseudo relevance feedback (PRF), which are used to 
provide the query-specific context into the model.
% Together with the origin query,
% these $m$ selected documents are used to evaluate the relevance of individual candidate documents.
Recall that, when using cross attention to model the relevance 
with BERT, the token sequence from a query $q$ and from a document $d$
are first concatenated into $[CLS]Query[SEP]Document[SEP]$
before passing through multiple self-attention layers, and 
the interaction representation for $[CLS]$ is used to denote the relevance between the query and the document~\cite{DBLP:conf/naacl/DevlinCLT19}.
In the calibration building block, intuitively, these $m$ PRF documents are selected to 
calibrate the modeling of the interaction between the query and individual documents, providing query-specific information from the PRF documents. When evaluating the relevance a document,
instead of directly using the interaction presentation $[CLS]$ between the token sequence from the document and the query,
we first use the $m$ prototypes to calibrate it. 
During the groupwise scorer, inspired by~\cite{DBLP:conf/ictir/AiWBGBN19, DBLP:conf/sigir/PangXALCW20},
instead of independently evaluating the relevance of individual documents,
$n$ documents are considered together using a four-layers BERT model,
and the relevance of these $n$ documents are evaluated jointly. 
% The relevance of $d$ is based on the calibrated interaction representation.
Put the PRF calibrator and the groupwise scorer together, 
the proposed Co-BERT is formulated as $Co\text{-}BERT(q, \mathcal{R}_m, d; \Theta)$, where $\mathcal{R}_m$
denotes the PRF document set with the $m$ selected documents and 
$\Theta$ is the trainable parameters. 
During training, the $\Theta$ is optimized to make better uses of the
$m$ feedback documents for the calibration, scoring individual documents using the groupwise scorer more effectively.
During inference, given a query $q$, 
% after selecting $m$ documents $\mathcal{R}_m$ (we still use the pre-trained model from MS-MARCO),
we score individual documents $d$ as in Eq.~\ref{eq.bert_prf}.
\begin{equation}\label{eq.bert_prf}
    \mathit{rel}(q, d) = Co\text{-}BERT(q, \mathcal{R}_m, d; \Theta)
\end{equation}
In both training and inference, given a query, we first need to 
select $m$ documents using the pre-trained model on MS MARCO~\cite{nguyen2016ms,DBLP:journals/corr/abs-1901-04085}.
The architecture of Co-BERT can be summarised into the following two components,
which could be trained in an end-to-end manner. 
\begin{enumerate}
%     \item  Construct $m$ interaction prototypes relative to the query using BERT, one for each  
% $m$ selected documents;
\item Given a document, we first get its interaction presentation 
relative to the query, and calibrate the representation using $m$ prototypes based on the $m$ selected PRF documents;
\item The relevance score of the document is determined jointly using groupwise scorer based on BERT, taking cross-documents interaction into consideration.
\end{enumerate}

Likewise in~\cite{DBLP:conf/sigir/DaiC19}, since a document could be too long to be encoded using BERT model, 
we split a document into overlapped passages with the same length. 
Similar to BERT-QE~\cite{zheng2020bert}, a BERT checkpoint pre-trained on MS MARCO~\citep{DBLP:journals/corr/abs-1901-04085} 
is used to score each passage relative to the query, and 
the passage with the highest score in each document is actually used 
in place of the original document in both training and inference.
For brevity, we use the term document in the following.

\subsection{Interaction Calibration using Pseudo Relevance Feedback}\label{sec.calibrate_q}
\noindent\textbf{Construct $m$ prototype interaction representations.}
We first construct 
prototypes for the interaction representations
using the $m$ selected PRF documents.
In particular, 
akin to the prior work~\citep{DBLP:journals/corr/abs-1901-04085}, 
the query and each of these PRF documents
are first concatenated into sequence $[CLS]Query[SEP]Doc[SEP]$, and is 
encoded using multiple BERT layers.
The $m$ output embedding of the token $[CLS]$ from BERT, each for one of 
the PRF documents, encode the interaction between the query and the corresponding PRF documents.
We denote these $m$ $[CLS]$ vectors as $t_i$, where $i\in [1,\cdots ,m]$,
and $t_i$ is a dense vector in $l$-dimension, e.g., $l=768$ if using BERT-Base.
A BERT-Base model is used in this work to generate the prototypes, which
is initialised with the pre-trained relevance model on MS Marco~\cite{DBLP:journals/corr/abs-1901-04085}. 

\noindent\textbf{Calibrate the interaction representations of the candidate documents.}
% \subsection{Calibrate the Interaction Presentation using Prototypes}\label{sec.encode_doc}
For the top-$k$ documents to be evaluated, 
similar to the vector of $t_i$,
we first employ the $[CLS]$ vector
after passing the concatenated $[CLS]Query[SEP]Doc[SEP]$
into multiple BERT layers, e.g., 12 layers if using BERT-base.
We denote these $k$ interaction vectors as $r_j$ where $j\in [1\cdots k]$, each corresponds to one document to be evaluated.
Thereafter, these $k$ vectors are calibrated 
using the $m$ prototypes $t_i$ with 
a shallow BERT model including two layers.
In particular, the interaction prototype $t_i$ and 
each interaction representation $r_j$ are stacked into a sequence with two tokens, namely, $t_i{r_j}$, before
passing through the two-layer BERT. 
The calibrated interaction representation corresponding to $r_j$
using prototype $t_i$
from the two-layers BERT output sequence is denoted as $rt_{ij}$.
Thereby, for each $r_j$, there are $m$ calibrated representations. 
Ultimately, we combine these $m$ calibrated presentations
into one using a simple weighted average, where the weight
is the relevance of the prototype $t_i$, as in Eq.~\ref{eq.calibrate_query}, where $W_t$ and $b_t$ are trainable weights for the projection.
Similar to the residual connection in the multi-head attention~\cite{DBLP:conf/nips/VaswaniSPUJGKP17}, 
as shown in Eq.~\ref{eq.calibrate_query_average},
we average the calibrated interaction representation and the origin presentation 
and use the resulting vector as the inputs for the follow-up scorer.
We show that this residual connection is important to the effectiveness in Section~\ref{sec.ablation}.
In this work, for the two-layers BERT model in the calibration,
we employ the configuration named \textit{uncased\_L-2\_H-768\_A-12}, which is with two layers, hidden size equaling 768, and 12 attention heads. 
We use the pre-trained BERT checkpoint from Google\footnote{\url{https://storage.googleapis.com/bert_models/2020_02_20/uncased_L-2_H-768_A-12.zip}}
to initialise this model. 

\begin{equation}\label{eq.calibrate_query}
     r_j' = \sum_{i\in [1\cdot m]}\operatorname{softmax}(W_t t_i + b_t) \cdot rt_{ij}
\end{equation}

\begin{equation}\label{eq.calibrate_query_average}
     \hat{r}_j = \frac{r_j + r_j^\prime}{2}
\end{equation}

\subsection{Groupwise Scorer using BERT}\label{sec.group_relevance}
With the calibrated interaction representation $\hat{r_j}$ for each of the $k$ documents,
in this section, we score individual documents using a novel groupwise scorer based on BERT. 
Intuitively, as mentioned in~\cite{DBLP:conf/ictir/AiWBGBN19},
scoring individual documents independently could lead to
sub-optimal ranker due to the comparing natural in the ranking problem.
Inspired by~\cite{DBLP:conf/ictir/AiWBGBN19,DBLP:conf/sigir/PangXALCW20}, we propose a groupwise relevance scorer using BERT, 
hoping to evaluate the document relevance more effectively 
by encoding the cross-documents interactions from the same ranking.

\textbf{The BERT-based groupwise scorer.}
In particular, instead of evaluating their relevance independently,
we group $n$ candidate documents ($n\le k$) together,
% maintaining their relative positions in the initial ranking, 
before modeling the relevance of these $n$ documents jointly.
Recall that transformer~\cite{DBLP:conf/nips/VaswaniSPUJGKP17} model relies on 
the positional embedding to encode the position information.
According to our pilot experiments, 
we do not configure the positional embedding within a group, and
simply generate different groups following the initial ranking.
For example, the first group 
includes documents ranked between one and ten, 
and the second group includes documents ranked between nine and eighteen, and so on 
(the overlap between groups will be explained in the following).
% position information 
For the $n$ documents in a single group, their 
interaction representations, e.g., $\hat{r_.}$
from Eq.~\ref{eq.calibrate_query_average},
are stacked into a sequence with length $n$, e.g., $\hat{r}_{1}\hat{r}_{2}\hat{r}_{3}\cdots \hat{r}_{n}$.
Thereafter, this sequence of interaction representations 
are passed through multiple layers of BERT, before being projected into
$n$ relevance scores, which are used to rank the documents.
Herein, a four-layer BERT model named \textit{uncased\_L-4\_H-768\_A-12}, with four layers, 
the hidden size of 768, and, 12 attention heads, is used.
We initialise this four-layers BERT model 
using pre-trained checkpoint from Google\footnote{\url{https://storage.googleapis.com/bert_models/2020_02_20/uncased_L-4_H-768_A-12.zip}}.
The choice of $n$ is up to the maximum batch size that is allowed
by the hardware.

\noindent\textbf{Overlapping between the neighbouring groups.}
As discussed, the cross-documents interaction from the same ranking
is believed crucial for an effective ranking.
Due to the hardware limitation, however, we can not feed all $k$ documents into one group and have to
generate multiple groups when $k$ is large, e.g., $k=1000$ in this work. 
To maintain the cross-references among different groups, 
we employ a straightforward method by allowing an overlap with $o$ documents, e.g., $o=4$, 
in between neighbouring groups from the initial ranking. 
For example, when $k=1000$, $n=200$ and $o=5$, 
given these $d_i$ with $i\in[1,\cdots, 1000]$,
we will have  six groups with documents as follows. 
% $$[[d_1,\cdots, d_200], [d_196,\cdots, d_396], [d_392,\cdots, d_592], [d_588,\cdots, d_788], [d_784,\cdots, d_984], [d_980,\cdots, d_1000]]$$
$$[[d_1,\cdots, d_{200}], [d_{196},\cdots, d_{395}], \cdots, [d_{976},\cdots, d_{1000}]]$$
The last group with less documents is zero padded to batch size.

\subsection{End-to-end Training of the Model}\label{sec.end2end_train}
As mentioned in Eq.~\ref{eq.bert_prf},
we could learn trainable weights denoted as $\Theta$ in
$Co\text{-}BERT(q, \mathcal{R}_m, d; \Theta)$.
The $\Theta$ includes the following parameters.
\begin{itemize}
    \item Trainable weights in the BERT-Base that generates the initial interaction representation, namely, $t_i$ and $r_j$,
    for the prototypes and individual documents in Section~\ref{sec.calibrate_q},
    which share the same weights;
    \item Trainable weights in the two-layer BERT  and 
    the $W_t$ and $b_t$ for projections
    in Eq.~\ref{eq.calibrate_query}, where
    all weights are shared among different calibrations using different 
    prototypes;
    \item The trainable weights in the four-layers 
    BERT used in groupwise scorer in Section~\ref{sec.group_relevance}.

\end{itemize}

Given a query $q$ and $k$ documents,
we first select $m$ PRF documents as described in
Section~\ref{sec.overview} using BERT ranker pre-trained on MS Marco~\cite{DBLP:journals/corr/abs-1901-04085}.
Thereafter, the batch size is determined based on the constrains of the GPU hardware. Therein, in each batch, 
$n$ candidate documents, together with the $m$ PRF documents are batched together.
During training, cross-entropy loss is computed 
for individual documents as in Eq.~\ref{eq.ce_loss},
where $I_{pos}$ and $I_{neg}$ denote the sets of indexes for relevant and non-relevant documents, respectively, and $pr_j$ is the probability of the document $j$ being relevant according to the model.
The probability is computed using a $\mathit{softmax}$ function, namely,
$pr_j=\mathit{softmax}(\mathit{rel}(q,d))$, where 
$\mathit{rel}(q,d)$ is based on Co-BERT as in Eq.~\ref{eq.bert_prf}.
  \begin{equation}\label{eq.ce_loss}
      \mathcal{L}(I_{pos}, I_{neg}, q, d_j) = - \sum_{j\in I_{pos}}{\log(pr_j)-\sum_{j\in I_{neg}}{\log(1-pr_j)}}
 \end{equation}
Note that, different from~\cite{DBLP:conf/ictir/AiWBGBN19, DBLP:conf/sigir/PangXALCW20},
we use pointwise loss as in Eq.~\ref{eq.ce_loss} to train the groupwise scorer, and
the cross-documents interaction is implemented using the two-layers BERT-based calibrator in Section~\ref{sec.calibrate_q}
and the four-layers BERT-based groupwise scorer described in Section~\ref{sec.group_relevance}. 
\section{Experiment Setup}\label{sec.exp_setup}

\begin{table*}[!t]
\centering
\caption{Effectiveness of Co-BERT relative to baseline models. The comparisons are mainly relative to the best-performed BERT-based model that is pre-trained on MS Marco before fine-tuning, denoted as BERT-Base (MS Marco). In addition, the results for the recent state-of-the-art PRF model using BERT, namely, BERT-QE~\cite{zheng2020bert}, is also included for comparisons. The relative gain/loss in terms of percentage (in the bracket) relative to BERT-Base (MS Marco) is reported. The statistical significance at 0.05 relative to BERT-Base (MS Marco) and BERT-QE are denoted as $\dagger$ and $\ddagger$,  respectively.}\label{tab.effectiveness} 
\resizebox{\textwidth}{!}{
\begin{tabular}{l|llllll}
\toprule
\multirow{2}{*}{Model} & \multicolumn{3}{c|}{Robust04}              & \multicolumn{3}{c}{Gov2}            \\ 
\cline{2-7}     
&   P@20   & NDCG@20  & \multicolumn{1}{l|}{MAP@1K} & P@20   & NDCG@20  & MAP@1K \\ 
\hline
BM25+RM3~\cite{DBLP:conf/ecir/LinCTCCFIMV16}      & 0.3821 & 0.4407    & \multicolumn{1}{l|}{0.2903} & 0.5634 & 0.4851    & 0.3350 \\
DPH+KL~\cite{DBLP:conf/sigir/MacdonaldMSO12} & 0.3924 & 0.4397    & \multicolumn{1}{l|}{0.3046} & 0.5896 & 0.5122    & 0.3605 \\
\hline
BERT-Base~\cite{DBLP:conf/sigir/DaiC19,padaki2020rethinking}       & 0.4070 & 0.4670    & \multicolumn{1}{l|}{0.2320} & - & -  & - \\
FilteredGoogleQuestions~\cite{padaki2020rethinking}   & 0.4130 & 0.4860   & \multicolumn{1}{l|}{0.2390} & - & -   & - \\
BERT-Base (MS Marco)      & 0.4430 & 0.5109   & \multicolumn{1}{l|}{0.3407} & 0.5725 & 0.5040   & 0.3531 \\
BERT-QE~\cite{zheng2020bert}       & 0.4602  & \textbf{0.5245}    & \multicolumn{1}{l|}{0.3571} & 0.5997  & 0.5206  & 0.3574 \\
\hline
Co-BERT (PRF calibrator only) 
            & 0.4526$^{\dagger}$ (+2.2\%)       & 0.5180 (+1.4\%)  & \multicolumn{1}{l|}{0.3480$^{\dagger}$(+2.1\%)} 
            & 0.5802 (+1.3\%)                        & 0.5026 (-0.3\%)
            & 0.3550 (+0.5\%) \\ 
% 2,2%	1,4%	2,1%	1,3%	-0,3%	0,5%
Co-BERT (groupwise scorer only)
            & 0.4500 (+1.6\%)           & 0.5101 (-0.2\%)
            & \multicolumn{1}{l|}{0.3530$^{\dagger}$ (+3.6\%)} 
            & 0.6493$^{\dagger\ddagger}$ (+13.4\%)               & 0.5640$^{\dagger\ddagger}$ (+11.9\%)    
            & 0.3993$^{\dagger\ddagger}$ (+13.1\%) \\
% 1,58%	-0,16%	3,61%	13,41%	11,90%	13,08%       
Co-BERT   & \textbf{0.4629}$^{\dagger}$ (+4.5\%) & 0.5213 (+2.0\%)    & \multicolumn{1}{l|}{\textbf{0.3631}$^{\dagger}$ (+6.6\%)}  & \textbf{0.6668}$^{\dagger\ddagger}$ (+16.5\%)& \textbf{0.5781}$^{\dagger\ddagger}$ (+14.7\%) &\textbf{0.4022$^{\dagger\ddagger}$} (+13.9\%) \\

\bottomrule
\end{tabular}}
\end{table*}

\begin{table*}[!h]
\centering
\caption{The computation cost of the models in comparisons in terms of FLOPs during inference.}
\label{tab.flops}
\resizebox{\textwidth}{!}{%
\begin{tabular}{l|lllllll}
\toprule
\multirow{3}{*}{Model} &
  \multicolumn{7}{c}{Robust04} \\ \cline{2-8} 
 &
  \multicolumn{1}{l|}{\multirow{2}{*}{$\sharp$Q}} &
  \multicolumn{2}{l|}{Re-rank the Initial Ranking} &
  \multicolumn{2}{l|}{Second Re-ranking using PRF} &
  \multicolumn{1}{l|}{\multirow{2}{*}{FLOPs(T)/Query}} &
  \multirow{2}{*}{total} \\ \cline{3-6}
 &
  \multicolumn{1}{l|}{} &
  $\sharp$Passages &
  \multicolumn{1}{l|}{FLOPs(T)} &
  $\sharp$Passages &
  \multicolumn{1}{l|}{FLOPs(T)} &
  \multicolumn{1}{l|}{} &
  \\ \hline
BERT-Base (MS Marco) &
  \multicolumn{1}{l|}{\multirow{5}{*}{249}} &
  2,729,808 &
  \multicolumn{1}{l|}{59,356.224} &
  0 &
  \multicolumn{1}{l|}{0.000} &
  \multicolumn{1}{l|}{238.378} &
  1.000x \\ \cline{1-1} \cline{3-8} 
BERT-QE \cite{zheng2020bert} &
  \multicolumn{1}{l|}{} &
  3.904,616 &
  \multicolumn{1}{l|}{255,739.021} &
  249,000 &
  \multicolumn{1}{l|}{182,868.894} &
  \multicolumn{1}{l|}{1,761.478} &
  7.389x \\ \cline{1-1} \cline{3-8} 
Co-BERT &
  \multicolumn{1}{l|}{} &
  \multirow{3}{*}{2,729,808} &
  \multicolumn{1}{l|}{\multirow{3}{*}{59,356.224}} &
  288,000 &
  \multicolumn{1}{l|}{6,301.095} &
  \multicolumn{1}{l|}{263.684} &
  1.106x \\
Co-BERT w/PRF calibrator only &
  \multicolumn{1}{l|}{} &
  &
  \multicolumn{1}{l|}{} &
  288,000 &
  \multicolumn{1}{l|}{6,293.448} &
  \multicolumn{1}{l|}{263.653} &
  1.106x \\
Co-BERT w/groupwise scorer only &
  \multicolumn{1}{l|}{} &
  &
  \multicolumn{1}{l|}{} &
  256,000 &
  \multicolumn{1}{l|}{5,594.176} &
  \multicolumn{1}{l|}{260.845} &
  1.094x \\ \hline
\multirow{3}{*}{Model} &
  \multicolumn{7}{c}{Gov2} \\ \cline{2-8} 
 &
  \multicolumn{1}{l|}{\multirow{2}{*}{$\sharp$Q}} &
  \multicolumn{2}{l|}{Re-rank the Initial Ranking} &
  \multicolumn{2}{l|}{Second Re-ranking using PRF} &
  \multicolumn{1}{l|}{\multirow{2}{*}{FLOPs(T)/Query}} &
  \multirow{2}{*}{total} \\ \cline{3-6}
 &
  \multicolumn{1}{l|}{} &
  $\sharp$Passages &
  \multicolumn{1}{l|}{FLOPs(T)} &
  $\sharp$Passages &
  \multicolumn{1}{l|}{FLOPs(T)} &
  \multicolumn{1}{l|}{} &
  \\ \hline
BERT-Base(MS Marco) &
  \multicolumn{1}{l|}{\multirow{5}{*}{150}} &
  7,375,231 &
  \multicolumn{1}{l|}{160,364.645} &
  0 &
  \multicolumn{1}{l|}{0.000} &
  \multicolumn{1}{l|}{1,069.098} &
  1.000x \\ \cline{1-1} \cline{3-8} 
BERT-QE \cite{zheng2020bert} &
  \multicolumn{1}{l|}{} &
  10,954,603 &
  \multicolumn{1}{l|}{717,489.106} &
  150,000 &
  \multicolumn{1}{l|}{109,941.457} &
  \multicolumn{1}{l|}{5,516.204} &
  5.160x \\ \cline{1-1} \cline{3-8} 
Co-BERT &
  \multicolumn{1}{l|}{} &
  \multirow{3}{*}{7,375,231} &
  \multicolumn{1}{l|}{\multirow{3}{*}{160,364.645}} &
  172,800 &
  \multicolumn{1}{l|}{3,780.657} &
  \multicolumn{1}{l|}{1,094.302} &
  1.024x \\
Co-BERT w/PRF calibrator only &
  \multicolumn{1}{l|}{} &
  &
  \multicolumn{1}{l|}{} &
  172,800 &
  \multicolumn{1}{l|}{3,776.069} &
  \multicolumn{1}{l|}{1,094.271} &
  1.024x \\
Co-BERT w/groupwise scorer only &
  \multicolumn{1}{l|}{} &
  &
  \multicolumn{1}{l|}{} &
  153,600 &
  \multicolumn{1}{l|}{3,356.505} &
  \multicolumn{1}{l|}{1,091.474} &
  1.021x \\ \bottomrule
\end{tabular}%
}
\end{table*}

\subsection{Dataset and Metrics}\label{sec.dataset_exp}
Likewise in~\cite{DBLP:conf/emnlp/YilmazYZL19,DBLP:conf/cikm/GuoFAC16}, we use the widely-used Robust04~\cite{DBLP:conf/trec/Voorhees04b} and GOV2~\cite{DBLP:conf/trec/ClarkeCS04} as test collections.
Robust04 consists of 528K documents and GOV2 consists of 25M documents. 
We employ 249 TREC keyword queries for Robust04 and 150 keyword queries for GOV2.
We report P@20, nDCG@20 to enable the comparisons on the shallow pool; and MAP@1K is reported for deep pool.
In addition, statistical significance for paired two-tailed t-test is reported.%at significant levels 0.01, 0.05, and 0.1 are reported.

\subsection{Initial Ranking and Baseline Models}\label{sec.baselines}

\noindent\textbf{DPH+KL} is used to generate the initial ranking with top-1k documents.
DPH is an unsupervised retrieval model derived from the divergence-from-randomness framework~\cite{DBLP:conf/trec/AmatiABGG07}. 
DPH+KL ranks the documents with DPH after expanding the original queries with Rocchio's query expansion using Kullback-Leibler divergence~\cite{DBLP:phd/ethos/Amati03,rocchio1971relevance}.
The implementation from Terrier toolkit~\cite{DBLP:conf/sigir/MacdonaldMSO12} has been adopted.
The performance of DPH+KL has also been reported for references. 

\noindent\textbf{BM25+RM3} is another unsupervised ranking model using pseudo relevance feedback signals,
wherein RM3~\cite{DBLP:conf/sigir/LavrenkoC01} is adopted to expand the queries.
We follow the experimental settings 
from~\cite{DBLP:conf/emnlp/YilmazYZL19}, and the implementation from
Anserini~\cite{DBLP:conf/ecir/LinCTCCFIMV16} with default settings is used. 

\noindent\textbf{BERT-Base}~\cite{DBLP:conf/sigir/DaiC19, padaki2020rethinking} is the vanilla BERT-Base model 
fine-tuned on the target dataset using up to top-30 text chunks of each relevant document as positive examples.
The MaxP variant is used for scoring individual documents, based on the passage with maximal relevance score.
The results from~\cite{DBLP:conf/sigir/DaiC19, padaki2020rethinking} are directly included.

\noindent\textbf{FilteredGoogleQuestions~\cite{padaki2020rethinking}} is the best-performed 
query expansion methods when using BERT re-ranker in~\cite{padaki2020rethinking}.
It expands the query
using Google’s
suggestions for reformulated queries to acquire additional related questions.
According to the authors, the best-performed data processing during the training and testing
has been used, namely,
the BERT model is
trained on descriptions from Robust04, and during evaluation,
a concatenation of the original
title-query with the extensions of the original query is used.
The results from~\cite{padaki2020rethinking} are directly included.

\noindent\textbf{BERT-Base (MS Marco)} is another ranking model using BERT-Base, and has been significantly
boosted using transfer learning. The model is initialised using a 
checkpoint that has been trained on MS Marco~\cite{DBLP:journals/corr/abs-1901-04085}, before being
fine-tuned on target datasets. 
As the model is already trained with large amount of data from MS Marco, on the target dataset, we only employ
the top-1 passage for each relevant document as positive examples during the fine-tuning.
This configuration is similar to BERT-QE~\cite{zheng2020bert}, which could dramatically reduce the 
training time without dampening the performance.

\noindent\textbf{BERT-QE}~\cite{zheng2020bert} is a recently proposed BERT-based re-ranking model exploiting the pseudo relevance feedback signals.
Unlike Co-BERT, BERT-QE is an inference framework and has not been trained end-to-end.
In this work, to enable comparisons, we use the BERT-QE variances using three BERT-Base components (namely, BERT-QE-BBB),
each for one of the phases. For the remaining, we follow the configurations from the original paper~\cite{zheng2020bert}, namely,
using sliding window including 100 words with an overlap of 50 words to generate the ``passages'' in MaxP, and 
the maximum length of the concatenated token sequence is set to 384.

\subsection{Variants of Co-BERT}\label{sec.bertref_var}

\noindent\textbf{Co-BERT} is the model as described in Section~\ref{sec.method} using BERT-based groupwise scorer
on top of the calibrated interaction representations based on PRF.

\noindent\textbf{Co-BERT with PRF calibrator only} is a variant of Co-BERT. 
Instead of using groupwise scorer, 
the relevance of documents are evaluated independently using
Eq.~\ref{eq.pointwise} without passing the batch of 
calibrated interaction representations into the BERT.

Without using the groupwise scorer effectively removes 
the cross-documents interaction from the model.
In particular, we simply project 
individual $\hat{r}_j$ from Eq.~\ref{eq.calibrate_query_average}
into a relevance score using a shared trainable weights $W_{rel}$ and $b_{rel}$ for each of the $k$ documents, as in Eq.~\ref{eq.pointwise}. 
\begin{equation}\label{eq.pointwise}
    rel(q, \mathcal{R}_m, d_j)=W_{rel}\hat{r}_j+b_{rel}
\end{equation}

\noindent\textbf{Co-BERT with group scorer only} is another variant of Co-BERT
without using the PRF calibration, and only use the groupwise scorer described in Section~\ref{sec.group_relevance}.
In particular, 
we turn off the PRF calibrator from Eq.~\ref{eq.calibrate_query} and~\ref{eq.calibrate_query_average}
and directly feed $r_j$ for document $j$ into the groupwise scorer from
Section~\ref{sec.group_relevance}. This means we do not use any feedback signals in the re-ranking,
but still use the groupwise scorer for training and inference.

Note that the results for the baselines and the Co-BERT variants 
are based on the standalone ranking models
\textit{without} the interpolation with the unsupervised ranking score.

\subsection{Model Training and Inference}\label{sec.model_training_exp}
\noindent\textbf{Data preparation.}
Instead of using all of the annotated documents for training, both training and inference are based on the top-1000 documents from DPH+KL. 
% We train on the top 1,000 documents returned by the unsupervised DPH+KL baseline. 
Akin to~\cite{DBLP:conf/sigir/DaiC19},
for BERT-Base (MS Marco) and Co-BERT, 
the documents are chunked using sliding windows including 150 words and 
the neighbouring windows are overlapped with 75 words.
As mentioned in Section~\ref{sec.overview}, for both Co-BERT and BERT-Base,
the most relevant passage is selected using a pre-trained BERT ranker on MS Marco~\cite{DBLP:journals/corr/abs-1901-04085}
to represent individual documents.
To feed individual query-paragraph (namely, the text chunk with 150 words) pairs into the model, 
the query and the paragraph 
are concatenated and the maximum sequence length is set to 256.

\noindent\textbf{Batching and loss function.}
We train BERT-Base (MS Marco) and Co-BERT using cross-entropy loss as in Eq.~\ref{eq.ce_loss}
for five epochs with a batch size of 64 on one NVIDIA TITAN RTX 24G. For Co-BERT, according to preliminary experiments, 
we configure the number of PRF documents for calibration as four ($m=4$),  
the number of candidate documents in 
individual group as 60 ($n=60$), and the overlap between the neighbouring groups is set to four ($o=4$).
During training, we randomly shuffle the batches before feeding  them into the model. 
In Section~\ref{sec.analysis}, we will discuss the impacts of different configurations for these hyper-parameters. 
The Adam optimizer~\cite{DBLP:journals/corr/KingmaB14} is used with the learning rate schedule from~\cite{DBLP:journals/corr/abs-1901-04085}.
We configure the initial learning rate as 3e-6, and the warming up steps are set to the 10\% of the total training steps. 

\noindent\textbf{Cross-validation.}
Similar to the configuration in DRMM~\cite{DBLP:conf/cikm/GuoFAC16}, we use 5-fold cross-validation to report the results.
Namely,
test queries are split into five equal-sized partitions from these two test collections.
The query partition on Robust04 follows the settings from~\cite{DBLP:conf/sigir/DaiC19}. 
On GOV2, queries are partitioned by the order of TREC query id in a round-robin manner. 
In each fold, three partitions are used for training, one is for validation, and the remaining one is for testing. 
In each fold, we conduct model selection
on the validation set and report the performance on test set based on the 
models with the highest nDCG@20 on the validation set. 
The ultimate performance is the average performance on the test splits from all folds.

\subsection{Computation of FLOPs}\label{sec.flops}

Akin to literature~\cite{DBLP:conf/acl/LiuZWZDJ20}, we report FLOPs (floating point operations) in Section~\ref{sec.efficiency}
which measures the computational complexity of models. Similar to~\cite{DBLP:conf/sigir/KhattabZ20}, we report FLOPs that includes all computations when using BERT-Base (MS Marco), BERT-QE, and Co-BERT 
during inference.

\section{Results}\label{sec.results}
In this section, we report results for the proposed Co-BERT model and compare it
to the baseline models. We first examine the effectiveness and efficiency of Co-BERT 
relative to baseline models in Section~\ref{sec.effectiveness} and Section~\ref{sec.efficiency}, respectively; 
Thereafter, in Section~\ref{sec.ablation}, we conduct ablation experiments 
to examine the impacts of different building blocks described in Section~\ref{sec.method}.

\subsection{Effectiveness of Co-BERT}\label{sec.effectiveness}

Given a query and top-1000 documents from DPH+KL, 
different BERT-based ranking models, including the multiple variants of Co-BERT model described in Section~\ref{sec.exp_setup}, are used to re-rank the 1000 documents.
We also include two state-of-the-art unsupervised ranking models, namely, BM25+RM3 and DPH+KL, for references. 
The ranking effectiveness are summarised on both shallow (using P@20 and nDCG@20) and deep pool (MAP@1K) in Table~\ref{tab.effectiveness}.

\noindent\textbf{Relative to the BERT-based re-ranker,}
the significant test as well as relative comparison in terms of percentage (in bracket)
in comparison with BERT-base (MS Marco)
are reported.
Co-BERT could outperform comfortably
the fine-tuned BERT-Base model from~\cite{DBLP:conf/sigir/DaiC19,padaki2020rethinking},
which does not use the pre-trained model from MS Marco~\cite{DBLP:journals/corr/abs-1901-04085}.
As mentioned in Section~\ref{sec.bertref_var}, both BERT-Base (MS Marco) and Co-BERT have been initialised 
using the relevance model pre-trained on MS Marco~\cite{DBLP:journals/corr/abs-1901-04085}, and are fine-tuned in a same way. 
Thereby, we are assured that the performance difference between Co-BERT and BERT-Base (MS Marco)
come from the novel model architecture introduced in Section~\ref{sec.method}.
As can be seen in Table~\ref{tab.effectiveness},
when only using either the PRF calibrator or the groupwise scorer, 
on Robust04, either variant can barely outperform the BERT-Base (MS Marco),
and the groupwise scorer only model even achieve slightly lower nDCG@20.
On GOV2, only the groupwise scorer can outperform the BERT-Base (MS Marco)
with a margin. 
When combining the two building blocks together,
Co-BERT actually outperforms both variants with single component.
Noticeably,
the complete Co-BERT could comfortably outperform BERT-Base (MS Marco) models on both Robust04 and Gov2 with a wide margin.
Actually, on the shallow pool, in terms of nDCG@20, 
Co-BERT improves upon BERT-Base (MS Marco) by 2.0\% and 14.7\% on Robust04 and Gov2, respectively.
On the deep pool, even bigger boosts have been observed, where Co-BERT improves 
BERT-Base by 6.6\% and 13.9\% on Robust04 and Gov2, respectively.
This confirms the effectiveness of the proposed model architecture, especially on deep pool.

\noindent\textbf{Comparisons relative to ranking models using pseudo relevance feedback signals}.
Due to the uses of PRF documents for calibration as described in Section~\ref{sec.calibrate_q},
we further include ranking models that also employ PRF signals for comparisons. 
According to Table~\ref{tab.effectiveness},
all three variants of Co-BERT could outperform the unsupervised PRF models, namely, BM25+RM3 and DPH+KL,
and, the fine-tuned BERT-Base model using query expansion from~\cite{padaki2020rethinking}
with a wide margin.  
In additions, the comparisons relative to the most recent BERT-based PRF model, namely, BERT-QE~\cite{zheng2020bert}, 
also confirm the superior effectiveness of the complete Co-BERT.
Similar to the observations relative to BERT-Base (MS Marco), however,
the PRF calibrator and groupwise scorer, 
when using independently, can not always outperform the BERT-QE actually.
For example, on Robust04, in terms of nDCG@20, 
the Co-BERT (PRF calibrator opnly) underperform BERT-QE by -1.2\%, 
and Co-BERT (groupwise scorer only) by -2.7\%.

When jointly using them together, 
on Robust04, Co-BERT performs on par with BERT-QE on shallow pool, meanwhile 
outperforms BERT-QE by 1.7\% in deep pool.
On Gov2, remarkably, Co-BERT could outperform the recent BERT-QE by a wide margin, for which we observe more than 10\% gains relative to this strong baseline on both shallow and deep pools.

\subsection{Efficiency of Co-BERT}\label{sec.efficiency}
Though we mainly focus on boosting the effectiveness in this work, 
we also report the efficiency of the proposed Co-BERT, comparing it with the baseline models. 
Similar to the established PRF models~\cite{DBLP:phd/ethos/Amati03,rocchio1971relevance, zheng2020bert}, 
Co-BERT needs to score the documents twice. Therein,
as described in Section~\ref{sec.overview},
we first score individual query-document pairs
to select $m$ documents 
as prototypes for the PRF calibration; thereafter, we re-rank the documents again using pseudo relevance feedback signals.
As described in Section~\ref{sec.overview}, however,
due to the uses of MaxP~\cite{DBLP:conf/sigir/DaiC19}, 
only in the first-round scoring,
all query-paragraph pairs are actually scored, resulting in $k\times \bar{c}$ forward pass using the deep ranking model, e.g., BERT-Base, where
$\bar{c}$ is the average number of paragraphs among the $k$ documents.  
When re-ranking again using pseudo relevance feedback, 
we only need to re-score the most relevant paragraph in each document,  
leading to $k$ forward pass in total for $k$ documents. Thereby, comparing with BERT-Base, the computation overhead is not really doubled. 
The FLOPs for different variants of Co-BERT, BERT-QE,  and BERT-Base(MS Marco) are reported in Table~\ref{tab.flops}.
Likewise in~\cite{zheng2020bert}, the FLOPs are presented in terms of the relative comparisons to the FLOPs when using BERT-Base.
From Table~\ref{tab.effectiveness} and~\ref{tab.flops}, comparing with BERT-Base(MS Marco),
it can be seen that
Co-BERT only requires extra 10.6\% computation overheads when significantly boosting the effectiveness 
on both shallow (2.0\%) and deep pool (6.6\%) on Robust04; meanwhile, with only 2.4\% extra computation cost, 
Co-BERT could provide more than 10\% boosts on both shallow and deep pool on GOV2.

Remarkably, though being able to outperform BERT-QE in most cases, 
Co-BERT only requires much less extra computation cost comparing with BERT-QE, making it 
an efficient and effective ranking model.
\section{Analysis}\label{sec.analysis}
\begin{table*}[!t]
\centering
\caption{Results for the analysis of  Co-BERT.
The impacts of the residual connections introduced in Eq.~\ref{eq.calibrate_query_average}
are studied in an ablation test.
In addition, 
two alternative feeding order of batches during training
are also investigated. 
Relative comparison in terms of percentage (in bracket)
in comparisons with BERT-Base (MS Marco) is also reported.
Statistical significance at levels 0.05 
is denoted with $\dagger$
 and $\ddagger$, relative
to BERT-base (MS Marco) and Co-BERT, 
respectively.}\label{tab.ablation} 
\resizebox{\textwidth}{!}{
\begin{tabular}{l|llllll}
\toprule
\multirow{2}{*}{Model}  & \multicolumn{3}{c}{Robust04}&\multicolumn{3}{c}{Gov2}            \\ 
\cline{2-7}     
&  P@20   & NDCG@20 & \multicolumn{1}{l|}{MAP@1K} & P@20   & NDCG@20 & MAP@1K \\ 
\hline

BERT-Base (MS Marco)        & 0.4430 & 0.5109  & \multicolumn{1}{l|}{0.3407} & 0.5725 & 0.5040   & 0.3531 \\

Co-BERT          & \textbf{0.4629} & \textbf{0.5213} & \multicolumn{1}{l|}{\textbf{0.3631}} 
                 & \textbf{0.6668} & \textbf{0.5781}  & \textbf{0.4022} \\ 
\hline

Co-BERT w/o residual connection in Eq.~\ref{eq.calibrate_query_average} 
                 & 0.4554$^{\dagger\ddagger}$ (-1.6\%) & 0.5102$^{\ddagger}$ (-2.1\%)  & \multicolumn{1}{l|}{0.3567$^{\dagger}$ (-1.8\%)} 
                 & 0.6326$^{\dagger\ddagger}$ (-5.1\%) & 0.5484$^{\dagger\ddagger}$ (-5.1\%)  & 0.3951$^{\dagger}$ (-1.8\%) \\ 
% -1,6%	-2,1%	-1,8%	-5,1%	-5,1%	-1,8%
\hline
Co-BERT (Train following initial ranking)   
                 & 0.4422$^{\ddagger}$ (-4.5\%) & 0.5029$^{\ddagger}$ (-3.5\%) & \multicolumn{1}{l|}{0.3457$^{\ddagger}$ (-4.8\%)} 
                 & 0.6211$^{\dagger\ddagger}$ (-6.9\%) & 0.5308$^{\dagger\ddagger}$ (-8.2\%) & 0.3728$^{\dagger\ddagger}$ (-7.3\%) \\ 
% -4,5%	-3,5%	-4,8%	-6,9%	-8,2%	-7,3%
Co-BERT  (Train reversing initial ranking) 
                 & 0.4454$^{\ddagger}$ (-3.8\%) & 0.5026$^{\ddagger}$ (-3.6\%)  & \multicolumn{1}{l|}{0.3429$^{\ddagger}$ (-5.6\%)} 
                 & 0.6322$^{\dagger\ddagger}$ (-5.2\%) & 0.5429$^{\dagger\ddagger}$ (-6.1\%)  & 0.3799$^{\dagger\ddagger}$ (-5.5\%) \\

\bottomrule

\end{tabular}}
\end{table*}
\begin{table*}
%考虑超参影响，包括batchsize，top文档数目，batch之间文档的overlap数目，attention layer，sequence layer数目
% k:batch size, k1:top_doc_num, o:overlap_num, la:attention layer num, ls:sequence layer num
\caption{Different hyper-parameter configurations of Co-BERT. Statistically significant differences relative to BERT-Base (MS Marco) at levels 0.01, 0.05, and 0.1 are denoted as ***, **, and *, respectively. 
Three hyper-parameters including the batch size, the number of prototype documents $m$, and 
the number of overlapped documents between batches $o$ are studied.
% k$_{1}$: \# of PRF documents. o: \# of overlapped documents between batches. 
%l$_{a}$: \# of PRF attention layers. l$_{s}$: \# of scoring layers. 
}\label{tab.hyper-parameters} 
\centering
\resizebox{0.8\textwidth}{!}{
\begin{tabular}{l|l|l|l|llllll}
\toprule
 & \multirow{2}{*}{\shortstack[c]{batch\\size}} & \multirow{2}{*}{$m$} & \multirow{2}{*}{o} & 
                 \multicolumn{3}{c}{Robust04}      & \multicolumn{3}{c}{Gov2}            \\ 
\cline{5-10} 
&  &  &  & P@20   & NDCG@20  & \multicolumn{1}{l|}{MAP@1K} & P@20   & NDCG@20  & MAP@1K \\
\hline
BERT-Base (MS Marco)     & 64  & -  & -   & 0.4430 & 0.5109    & \multicolumn{1}{l|}{0.3407} & 0.5725 & 0.5040    & 0.3531 \\
           
\hline
\multirow{3}{*}{Impacts of batch size}     & $\underline{32}$  & 4  & 4   & 0.4574$^{***}$ & 0.5080    & \multicolumn{1}{l|}{0.3531$^{***}$} & 0.6466$^{***}$ & 0.5651$^{***}$    & 0.4017$^{***}$ \\
      & $\underline{48}$  & 4  & 4   & 0.4556$^{***}$ & 0.5081    & \multicolumn{1}{l|}{0.3548$^{***}$} & 0.6523$^{***}$ & 0.5693$^{***}$    & 0.4033$^{***}$ \\
        & $\underline{64}$  & 4  & 4   & 0.4629$^{***}$ & 0.5213   & \multicolumn{1}{l|}{0.3631$^{***}$} & \textbf{0.6668$^{***}$} & 0.5781$^{***}$ & 0.4022$^{***}$ \\

%top doc: 3 & 5 & 6
\hline
\multirow{4}{*}{Impacts of $m$}     
      & 64  & $\underline{3}$  & 4   & 0.4494 & 0.5098  & \multicolumn{1}{l|}{0.3551$^{***}$} & 0.6530$^{***}$ & 0.5713$^{***}$  & 0.4012$^{***}$ \\
      & 64  & $\underline{4}$  & 4  & 0.4629$^{***}$ & 0.5213   & \multicolumn{1}{l|}{0.3631$^{***}$} & \textbf{0.6668$^{***}$} & 0.5781$^{***}$ & 0.4022$^{***}$\\
      & 64  & $\underline{5}$  & 4   & 0.4592$^{***}$ & 0.5151    & \multicolumn{1}{l|}{0.3561$^{***}$} & 0.6587$^{***}$ & \textbf{0.5785$^{***}$}    & \textbf{0.4075$^{***}$} \\
      & 64  & $\underline{6}$  & 4  & 0.4637$^{***}$ &  0.5244$^{**}$    & \multicolumn{1}{l|}{ 0.3632$^{***}$} & 0.6470$^{***}$ & 0.5589$^{***}$    & 0.3998$^{***}$ \\
\hline
\multirow{6}{*}{Impacts of $o$}       & 64  & 4  &  $\underline{0}$   & 0.4651$^{***}$ & 0.5233$^{**}$    & \multicolumn{1}{l|}{0.3604$^{***}$} & 0.6503$^{***}$ & 0.5660$^{***}$    & 0.4038$^{***}$ \\
      & 64  & 4  &  $\underline{1}$   & 0.4629$^{***}$ & 0.5195    & \multicolumn{1}{l|}{0.3603$^{***}$} & 0.6426$^{***}$ & 0.5635$^{***}$      & 0.3975$^{***}$ \\
      & 64  & 4  &  $\underline{2}$  & 0.4645$^{***}$ & 0.5234$^{**}$    & \multicolumn{1}{l|}{0.3614$^{***}$} & 0.6557$^{***}$ & 0.5704$^{***}$   & 0.4032$^{***}$ \\
      & 64  & 4  &  $\underline{3}$  & 0.4594$^{***}$ & 0.5187    & \multicolumn{1}{l|}{0.3586$^{***}$} & 0.6624$^{***}$ & 0.5717$^{***}$    & 0.4021$^{***}$ \\
      &64  & 4  &  $\underline{4} $   & 0.4629$^{***}$ & 0.5213   & \multicolumn{1}{l|}{0.3631$^{***}$} & \textbf{0.6668$^{***}$} & 0.5781$^{***}$ & 0.4022$^{***}$ \\
      & 64  & 4  & $\underline{5} $    & 0.4645$^{***}$ & 0.5172    & \multicolumn{1}{l|}{0.3592$^{***}$} & 0.6547$^{***}$ & 0.5687$^{***}$    & 0.4016$^{***}$ \\

\bottomrule
\end{tabular}}

\end{table*}
\subsection{The Impacts of Residual Connection}\label{sec.ablation}
The results have been summarised in Table~\ref{tab.ablation}.
% \noindent\textbf{Remove the averaging operation in Eq.~\ref{eq.calibrate_query_average}.}
As described in Section~\ref{sec.calibrate_q}, 
the calibration of the interaction representation in Eq.~\ref{eq.calibrate_query} 
exploits a shallow BERT model with two layers (\textit{uncased\_L-2\_H-768\_A-12}).
The averaging operation in Eq.~\ref{eq.calibrate_query_average} 
adds back the origin interaction representation, providing 
more direct connections between early layers and the scorer layers.
We argue that it actually functions similar to
the residual connections as in BERT~\cite{DBLP:conf/naacl/DevlinCLT19}.
From Table~\ref{sec.ablation},
it can be seen that, without the averaging operation,
the performances of Co-BERT drops significantly especially on shallow pool.
Noticeably, on GOV2, the performance is dampen more 
when using calibration without using the average operation than removing the calibration
all together (the Co-BERT with groupwise scorer only in Table~\ref{tab.effectiveness}).
% namely, the Co-BERT with groupwise scorer only, 
This highlights the importance to add this skip connection
after calibrating the interaction representation using pseudo relevance feedback.

\subsection{Feeding Order of the Training Data for Co-BERT}
As mentioned in Section~\ref{sec.end2end_train} and~\ref{sec.model_training_exp},
when the total number of documents for ranking (namely, $k$) is too large to be fed into single batch, we have to 
group $n < k$ documents into batches during training and inference.
However, as discussed in Section~\ref{sec.group_relevance},
Co-BERT actually relies on the cross-documents interactions among candidate documents for scoring,
which might require extra dependency among batches especially during training. 
As mentioned in Section~\ref{sec.model_training_exp},
we feed the data for training after random shuffling. 
In this section, we investigate two alternative ways for the feeding order of training data,
which are listed in the following.
Note that, among different epochs, the training data is still shuffled 
among queries to avoid over-fitting. 
\begin{itemize}
    \item \textbf{Training Co-BERT by feeding training samples following the order in initial ranking.}
    Herein, when feeding training batches for the same query, the batches
    are ordered following the initial ranking. For example, 
    the batch, which includes the documents ranked between 57 to 117 in the initial ranking,
    is followed by a batch with documents ranked between 114 and 174.
    The batches for different queries are shuffled.
    We use the same hyper-parameters as described in Section~\ref{sec.model_training_exp} by using $m=4$, $n=60$, and $o=4$.
    \item \textbf{Training Co-BERT by feeding training samples following the reversed order in initial ranking.}
    Likewise in the above configuration, 
    we follow certain order to feed in training batches. Differently, 
    the batches are fed in the reversed order of the initial ranking. 
    The other settings remain the same as above.
\end{itemize}
According to the results in Table~\ref{tab.ablation}, it can be seen that,
with the alternative feeding order for the training data, Co-BERT could still outperform BERT-Base on GOV2.
Such alternative order, however, leads to at least 3.5\% drops among all different metrics on both dataset and the resulting models
are significantly worth than Co-BERT trained using fully shuffled batches. 
% the order of training data does have non-trivial impact on the model performance. 
Noticeably, feeding training batches in reversed order seems to provide better results on GOV2.
% in general has better effectiveness than BERT-PRF$_{sequential}$, particularly on GOV2.
A possible explanation is that the model begins the learning from relatively ``easy'' samples, i.e. the bottom-ranked documents that can be easily recognized as being non-relevant, and gradually to the more difficult samples where the relevant and non-relevant documents share common keywords. For more detailed investigation of the optimised feeding order of the training data, we leave it for the future works.

\subsection{Hyper-parameter Tuning}
In this section, we study the impacts of several hyper-parameters introduced in Section~\ref{sec.method}.
The hyper-parameters studied include the batch size,
the number of prototype documents $m$, the number of overlapped documents among batches
$o$, which are introduced in Section~\ref{sec.group_relevance}.
% Different configurations of these hyper-parameters could affect the robustness of the proposed model. 
We tune these hyper-parameters to study their impacts on the retrieval performance, and report the obtained results in Table~\ref{tab.hyper-parameters}. 

\noindent\textbf{Using different configurations for batch size.}
From Table~\ref{tab.hyper-parameters}, it can be seen that 
the configuration of batch size could influence the performance.
The choice of batch size equalling 64 actually achieves the optimized 
performance on both datasets, and is actually the largest possible batch size given 
the hardware. 

\noindent\textbf{Number of prototype documents ($m$).}
Table~\ref{tab.hyper-parameters} shows that 
different configurations of $m$ do not lead to 
big differences in the effectiveness. In particular,
 configurations larger than $m=4$, namely, $m=5$ or $6$,
do not lead to significant drops or improvements of the effectiveness.
In the meantime, when using less than four prototype documents (e.g., $m=3$), 
the effectiveness drops significantly especially on Robust04. Actually,
when using $m=3$, Co-BERT achieve lower nDCG@20 than BERT-Base (MS Marco).
This actually emphasizes the effectiveness of 
the proposed PRF calibrator building block again, namely,
without (groupwise scorer only in Table~\ref{tab.effectiveness}) or with too few 
prototype documents could both hurt the performances.

\noindent\textbf{Number of overlapped documents ($o$) between batches.}
Finally, we experiment with different configurations of the number of overlapped 
documents between batches. 
As can be seen from Table~\ref{tab.hyper-parameters},
on Robust04, the model is robust with respect to the different choices of $o$
especially in terms of P@20 and MAP@1K. Whereas on GOV2, 
though Co-BERT could always outperform BERT-Base (MS Marco) using different $o$,
it can be seen that non-zero $o$ could contribute to better effectiveness especially on shallow pool.
\section{Conclusion}\label{sec.conclusion}

In this paper, we show that the relevance weighting of individual documents using BERT 
can be improved by considering both
the neighbouring documents from the same ranking list and the query-specific characteristics.
In particular, we propose an end-to-end BERT-based re-ranking models, named Co-BERT,
wherein PRF information is first used to calibrate the interaction representations
before the relevances of a group of documents are modeled jointly. 
Evaluation on two standard TREC test collections, namely, Robust04 and GOV2, 
    demonstrates that the proposed Co-BERT could advance the state-of-the-art BERT-based ranking model
    by a big margin. Namely, compared with fine-tuned BERT-Base ranker~\cite{DBLP:journals/corr/abs-1901-04085},
    Co-BERT could boost in terms of nDCG@20 and MAP@1K by 2.0\% and 6.6\%, 
    respectively, on Robust04; and improves 
    nDCG@20 and MAP@1K by 14.7\% and 13.9\%, 
    respectively, on GOV2.
    In addition,
    Co-BERT also significantly outperforms the recently-proposed
    BERT-QE model by a wide margin on GOV2 and perform on par with 
    BERT-QE on Robust04, but with way smaller computation overhead.
    
For the future works, we plan to investigate the incorporation of both local and 
query-specific context into different types of ranking models. In particular, we plan to investigate how to
incorporate the two kinds of context into the 
two-tower ranking models like DPR~\cite{karpukhin2020dense},  ColBERT~\cite{DBLP:conf/sigir/KhattabZ20}, 
and CoRT~\cite{wrzalik2020cort} to further boost their effectiveness;
In addition, for the content-context-aware lexicon matching models like DeepCT~\cite{dai2019context}, we believe
the incorporation of the context information in the ranking phases could also help;
Finally, for the memory-efficient 
transformers~\cite{tay2020long} that has been sparsified to encode long sequence, like Reformer~\cite{kitaev2019reformer},
Conformer~\cite{mitra2020conformer}, and, Longformer~\cite{beltagy2020longformer},
we plan to use them to replace the BERT model in Co-BERT, hoping to 
better incorporate the cross-documents interaction among all $k$ documents in one group (batch),
even when $k$ is very large, e.g., $k=1000$.


%\begin{acks}
%To Robert, for the bagels and explaining CMYK and color spaces.
%\end{acks}

\clearpage%\balancecolumns

\bibliographystyle{ACM-Reference-Format}
\bibliography{chen}

\appendix






\end{document}